\documentclass[10pt,twocolumn,letterpaper]{article}

\usepackage[pagenumbers]{cvpr} 
\definecolor{cvprblue}{rgb}{0.21,0.49,0.74}
\usepackage[pagebackref,breaklinks,colorlinks,allcolors=cvprblue]{hyperref}
\usepackage{xcolor}
\usepackage{colortbl}
\usepackage{booktabs}
\usepackage{multirow}
\usepackage{bbding}
\usepackage{adjustbox}
\usepackage{graphicx}   
\usepackage{siunitx}   
\def\paperID{7217} 
\def\confName{CVPR}
\def\confYear{2026}

\newcommand{\method}{\textbf{DualTAP}\xspace}
\newcommand{\methodnorm}{DualTAP\xspace}
\title{DualTAP: A Dual-Task Adversarial Protector for Mobile MLLM Agents}

\author{
  \textbf{Fuyao Zhang\textsuperscript{1}\quad
          Jiaming Zhang\textsuperscript{1}\thanks{Corresponding authors}\quad
          Che Wang\textsuperscript{1,2}\quad
          Xiongtao Sun\textsuperscript{1,3}\quad
          Yurong Hao\textsuperscript{1}} \\
  \textbf{Guowei Guan\textsuperscript{1} \quad
          Wenjie Li\textsuperscript{4} \quad
          Longtao Huang\textsuperscript{5} \quad
          Wei Yang Bryan Lim\textsuperscript{1}} \\
  \textsuperscript{1}Nanyang Technological University \quad
  \textsuperscript{2}Peking University \\
  \textsuperscript{3}Xidian University \quad
  \textsuperscript{4}Hebei Normal University \quad
  \textsuperscript{5}Alibaba Group \\
{\centering Project Page: \url{https://fyzhang1.github.io/DualTAP}}
}

\begin{document}
\maketitle
\begin{abstract}
The reliance of mobile GUI agents on Multimodal Large Language Models (MLLMs) introduces a severe privacy vulnerability: screenshots containing Personally Identifiable Information (PII) are often sent to untrusted, third-party routers. These routers can exploit their own MLLMs to mine this data, violating user privacy. Existing privacy perturbations fail the critical dual challenge of this scenario: protecting PII from the router's MLLM while simultaneously preserving task utility for the agent's MLLM.
To address this gap, we propose the \textbf{Dual-Task Adversarial Protector (DualTAP)}, a novel framework that, for the first time, explicitly decouples these conflicting objectives. DualTAP trains a lightweight generator using two key innovations: (i) a contrastive attention module that precisely identifies and targets only the PII-sensitive regions, and (ii) a dual-task adversarial objective that simultaneously minimizes a task-preservation loss (to maintain agent utility) and a privacy-interference loss (to suppress PII leakage). To facilitate this study, we introduce PrivScreen, a new dataset of annotated mobile screenshots designed specifically for this dual-task evaluation.
Comprehensive experiments on six diverse MLLMs (e.g., GPT-5) demonstrate DualTAP's state-of-the-art protection. It reduces the average privacy leakage rate by 31.6 percentage points (a 3.0$\times$ relative improvement) while, critically, maintaining an 80.8\% task success rate—a negligible drop from the 83.6\% unprotected baseline. DualTAP presents the first viable solution to the privacy-utility trade-off in mobile MLLM agents.

\end{abstract}

\section{Introduction}
\label{sec:intro}

\begin{figure}[ht]
    \centering
    \includegraphics[width=1.0\linewidth]{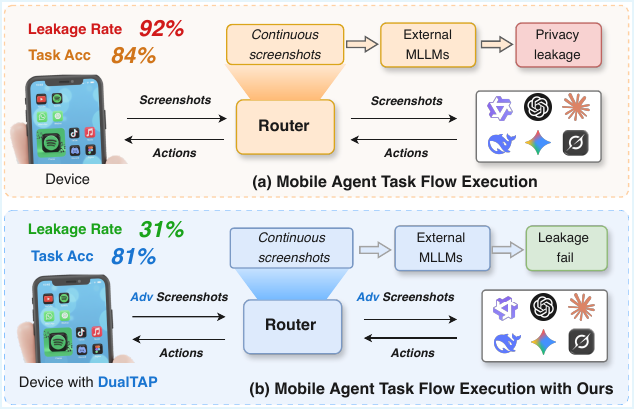} 
    \caption{The Dual Task Completion workflow of the GUI Agent, which is structured upon the Perceive-Router-MLLM paradigm. The figure presents two task flows: (a) In the absence of protection, the task flow achieves its functional objectives but inevitably leaks privacy tasks. (b) Deploying our method on the user side mitigates privacy leakage while preserving normal functionality.}
    \label{fig:mobile_agent_workflow}
\end{figure}

Multimodal large language models (MLLMs) are increasingly the core reasoning engines for Graphical User Interface (GUI) mobile agents~\cite{wang2024gui, ye2025mobile, wang2024mobile}.
Leveraging these models, such agents can handle a wide range of real-world tasks, including personal assistance, travel planning, and financial operations. As illustrated in Figure~\ref{fig:mobile_agent_workflow}~(a), these agents typically operate on a continuous Perceive–Router–MLLM loop. The Router acts as a centralized API scheduler, receiving user screenshots and task instructions and dispatching requests to a curated set of MLLM APIs to obtain an optimal response. This interaction paradigm, while facilitating coherent and efficient task completion~\cite{nguyen2024gui,yue2025masrouter}, introduces significant privacy risks.


The core vulnerability lies in the router, which we assume to be an honest-but-curious adversary. In a typical GUI agent scenario, a single user task is decomposed into multiple subtasks (Figure~\ref{fig:mobile_agent_workflow}~(a)), generating a large volume of sequential screenshots. The router gains access to this entire stream, allowing it to correlate information across continuous interactions. This creates a potent attack vector for the router provider: by leveraging its own MLLMs, it can automatically mine the screenshot stream to extract and reconstruct sensitive personally identifiable information (PII). Such unauthorized processing and profiling of personal data directly violates the core principles of the EU General Data Protection Regulation (GDPR)~\cite{european_commission_regulation_2016} and California Consumer Privacy Act (CCPA)~\cite{legislature2018california}, amplifying the privacy threat and exposing users to significant legal and reputational risks.



Existing privacy-preserving techniques often employ adversarial perturbations~\cite{zhang2025anyattack,jia2025adversarial,meftah2025vipvisualinformationprotection}, adding structured noise to inputs to disrupt an MLLM's ability to recognize sensitive content. However, these methods prove inadequate for the mobile agent scenario, as they fail to address the critical \emph{dual challenge}: (i) preserving \emph{task validity} for the agent's own MLLMs, while (ii) protecting \emph{user privacy} against the router's MLLMs. When enhancing privacy protection, most existing methods substantially compromise task utility.

To address this gap, we propose \textbf{Dual}-\textbf{T}ask \textbf{A}dversarial \textbf{P}rotector (\textbf{\method}), a pluggable module designed to protect PII within the mobile agent ecosystem (Figure~\ref{fig:mobile_agent_workflow}~(b)). Our generator first incorporates a contrastive attention module to precisely identify and target regions sensitive to privacy cues. We then optimize the generator using a dual-task adversarial objective, which simultaneously minimizes a task-preservation loss (ensuring agent utility) and maximizes a privacy-interference loss (suppressing sensitive information leakage). To facilitate this study, we introduce \textbf{PrivScreen}, a new benchmark dataset for evaluating privacy leakage in mobile MLLM agents. It contains over 500 high-resolution screenshots with synthetically injected PII, sourced from 10 common mobile applications across diverse usage scenarios.

We conduct a comprehensive evaluation of six state-of-the-art MLLMs, spanning open source (Qwen2.5, InternVL3), GUI specialized (Holo1.5, UI-TARS), and commercial systems (GPT-5, Gemini 2.0 Flash). \methodnorm provides robust protection: on our primary privacy leakage metric, it achieves an average reduction in leakage rate by 31.6 percentage points (a 3.0$\times$ relative improvement over the strongest baseline). Critically, it preserves utility with an 80.8\% task success rate, negligible degradation from the 83.6\% of the original, unprotected system. These results establish \methodnorm as the state-of-the-art for privacy protection in mobile MLLM agents.
Our contributions are three-fold: 
\begin{itemize} 
\item We propose \textbf{\method}, a novel dual-task adversarial framework that jointly optimizes for task preservation and privacy interference, effectively protecting sensitive information in screenshots without compromising agent performance. 
\item We introduce \textbf{PrivScreen}, a new dataset of 500+ privacy-sensitive screenshots with injected PII across 10 mobile apps, designed to evaluate and benchmark privacy leakage and protection mechanisms in GUI agents. 
\item We demonstrate through comprehensive experiments that \methodnorm achieves state-of-the-art privacy protection, significantly reducing information leakage while maintaining a high task success rate for the mobile agent. 
\end{itemize}

\section{Related Work}
\label{sec:formatting}


\paragraph{Mobile GUI Agents}
Mobile GUI agents have emerged as a research hotspot for automating complex tasks~\cite{wang2024mobile, hu2025agents, ye2025mobile}. Early approaches often relied on structured data or accessibility services. However, recent advances are dominated by MLLMs acting directly on visual input. These agents employ visual grounding to map natural-language instructions to specific UI elements~\cite{shao2024visual} or integrate it with chain-of-thought (CoT) reasoning to decompose intricate goals step-by-step~\cite{wei2022chain}. A standard workflow involves capturing device screenshots and routing them via APIs to MLLMs for contextual awareness and decision-making~\cite{yan2025mcpworld,guo2025mcp,jia2025osworldmcpbenchmarkingmcptool}. This paradigm enables end-to-end task execution without direct programmatic access to app internals~\cite{hu2024routerbench, zhang2025litewebagent}. Modern systems build on this by incorporating element-level metadata to enhance robustness~\cite{wang2024mobileagentbench,tang2025survey} or adopting CoT prompting to improve interpretability and mitigate errors in long-horizon tasks~\cite{wei2022chain,yang2025mla}.

\paragraph{PII Extraction by MLLMs} Early PII extraction relied on static scanning with keywords or regular expressions, enhanced by OCR~\cite{cui2025paddleocr30technicalreport} to parse screen data and match predefined sensitive patterns~\cite{kuang2021mmocrcomprehensivetoolboxtext, wei2024general, nazeem2024open}.
These methods depend on fixed signatures, lack contextual insight, and struggle to detect natural-language PII or link entities across screenshots for user profiling~\cite{mainetti2025detecting, kulkarni2021personally}.
Advanced MLLMs integrate visual encoders with language models, boosting the semantic understanding of on-screen content. State-of-the-art models like GPT-5~\cite{jaech2024openai} and Gemini-2.5~\cite{team2024gemini} interpret document semantics, UI details, and cross-frame associations, enabling comprehensive PII inference~\cite{yao2025efficient, wu2024visionllm} and uncovering hidden sensitive cues in complex interfaces.

\begin{figure*}[ht]
    \centering
    \includegraphics[width=1.0\linewidth]{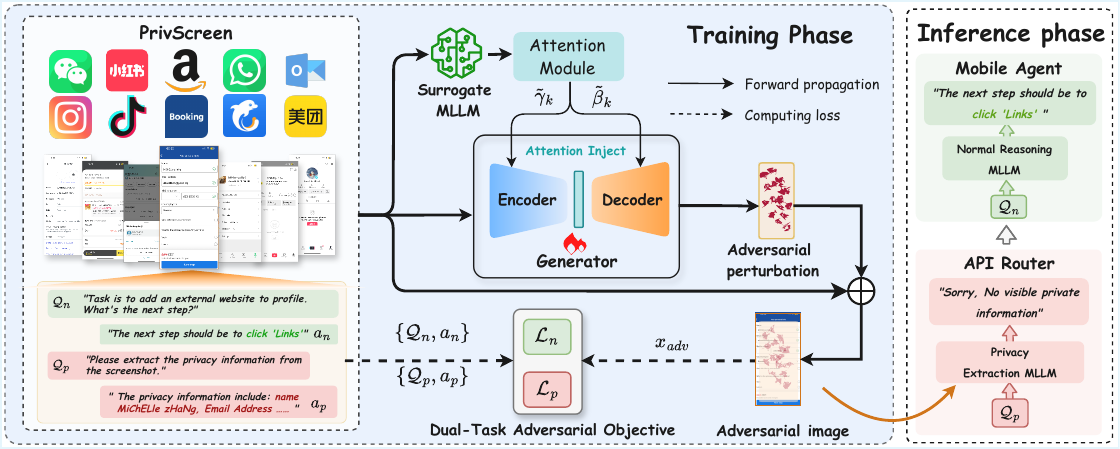}
    \caption{Overview of the proposed \method framework. \textbf{(a) Training Phase}: A perturbation generator $\mathcal{G}_\phi$ is trained using a dual-task objective. It learns to produce a perturbation $\delta$ based on a contrastive attention map $\mathcal{A}(x)$, guided by a surrogate MLLM $\mathcal{M}_\theta$. The objective is to simultaneously minimize the task-preservation loss $\mathcal{L}_{\text{n}}$ (for normal tasks $\mathcal{Q}_n$) and the privacy-interference loss $\mathcal{L}_{\text{p}}$ (for privacy tasks $\mathcal{Q}_p$). \textbf{(b) Inference Phase}: The pre-trained generator is deployed on the mobile device. It adds the optimized perturbation $\delta$ to the raw screenshot $x$ to create $x_{\mathrm{adv}}$. This $x_{\mathrm{adv}}$ is sent to the untrusted router, which blocks the router's Privacy Extraction MLLM ($\mathcal{M}_e$) from extracting PII, while still allowing the agent's Normal Task MLLM ($\mathcal{M}_a$) to function correctly.}
    \label{framework}
\end{figure*}

\paragraph{PII Protection via Adversarial Perturbations} As a promising privacy mechanism, adversarial perturbations inject structured noise into inputs to disrupt MLLM recognition~\cite{wang2025webinjectpromptinjectionattack, huang2025xtransferattackssupertransferable, wu2024improving}. Recent advances include Co-Attack~\cite{zhang2022towards}, which breaks image-text alignment via fusion-stage gradients; AttackVLM~\cite{zhao2023evaluating}, combining white-box proxy generation with black-box queries; AnyAttack~\cite{zhang2025anyattack}, using a self-supervised generator; FOA-Attack~\cite{jia2025adversarial}, enhancing transferability through global and local feature optimization; and VIP~\cite{meftah2025vipvisualinformationprotection}, which selectively perturbs regions of interest. Despite these advances, existing methods primarily focus on general tasks like image captioning or VQA. Their applicability to mobile agents is limited, as they are not designed for the specific dual-objective of selectively suppressing PII while preserving the UI-grounded task execution capabilities of the agent.

\section{Method}
\subsection{Preliminary}
\paragraph{Threat Model} 

We consider a privacy threat through the screenshot channel. The agent periodically captures screenshots $x$ and forwards them to remote MLLM APIs via a routing service (e.g., OpenRouter). We assume an honest-but-curious router, which faithfully relays requests but also analyzes the data it processes. The router employs a \emph{Privacy Extraction MLLM} $\mathcal{M}_e$ to analyze screenshots and extract sensitive PII for user profiling. It does not modify content or interfere with the agent's \emph{Normal Task MLLM} $\mathcal{M}_{a}$.


\paragraph{Problem Formulation} 

Our protector targets two black-box MLLMs: the router's $\mathcal{M}_e$ and the agent's $\mathcal{M}_a$. Both process the same input screenshot $x \in [0,1]^{3 \times H \times W}$. Given a privacy-related question-answer (QA) set $(\mathcal{Q}_p, a_p)$, $\mathcal{M}_e$ aims to extract the private text $a_p$. Conversely, for a normal task QA set $(\mathcal{Q}_n, a_n)$, $\mathcal{M}_a$ uses chain-of-thought reasoning $\mathcal{C} = (c_1, \ldots, c_k)$ to derive the benign answer $a_n$.
Our objective is to train a generator $\mathcal{G}_\phi$ that synthesizes an $L_\infty$-bounded adversarial perturbation $\delta$, yielding $x_{\mathrm{adv}} = \text{clip}_0^1(x + \delta)$ where $\|\delta\|_\infty \leq \varepsilon$. This achieves a dual goal: (i) disrupting $\mathcal{M}_e$'s privacy extraction, while (ii) preserving $\mathcal{M}_a$'s utility for normal tasks. Due to the black-box nature of $\mathcal{M}_e$ and $\mathcal{M}_a$, we employ a white-box surrogate model $\mathcal{M}_\theta$, which approximates the conditional distribution $p_\theta(a \mid q, x)$, to optimize $\mathcal{G}_\phi$. We rely on the transferability of perturbations from $\mathcal{M}_\theta$ to the target models.

\subsection{Dual-Task Adversarial Protector}
\paragraph{Overview}

As shown in Figure~\ref{framework}, \method~formulates the privacy-utility balance as a dual-objective optimization problem. We first introduce a contrastive attention module to generate a spatial map $\mathcal{A}(x)$ that isolates regions pertinent to privacy cues. This map is integrated into a U-Net-style generator $\mathcal{G}_\phi$ to guide the allocation of perturbations. Second, we optimize $\mathcal{G}_\phi$ using a dual-task adversarial objective, which explicitly trains the generator to trade off task usability and privacy security.

\paragraph{Contrastive Attention Module}
To isolate regions that are uniquely sensitive to privacy cues, we introduce an attention map $\mathcal{S}$. This map selectively highlights these responsive areas while occluding surrounding input regions, thereby distinguishing them from semantically salient features unrelated to privacy risks. In doing so, it enables precise, targeted generation of adversarial perturbations:

\begin{equation}
    \mathcal{S}(x;\mathcal{Q}) = \left|\frac{\partial \mathcal{L}_{\text{nll}}(x; \mathcal{Q})}{\partial x}\right| \in \mathbb{R}^{3 \times H \times W},
\end{equation}
where $\mathcal{L}_{\text{nll}}(x; \mathcal{Q}) = \sum_{(q,a) \in \mathcal{Q}} \big( -\log p_\theta(a \mid q, x) \big)$ is the total negative log-likelihood (NLL) loss over a QA set $\mathcal{Q}$, using the frozen surrogate model $\mathcal{M}_\theta$. This map quantifies the aggregate degradation in model confidence for each image $x$ from the dataset $\mathcal{D}$. 
To disentangle privacy-sensitive regions from task-salient regions, we propose a contrastive attention module:
\begin{equation}
\mathcal{A}(x) =\text{ReLU}(\mathcal{S}_{\mathrm{p}}(x)-\mathcal{S}_{\mathrm{n}}(x)),
\end{equation}
where $\mathcal{S}_{\mathrm{p}}(x) = \mathcal{S}(x; \mathcal{Q}_{\mathrm{p}})$ is the saliency map for privacy-oriented QA pairs and $\mathcal{S}_{\mathrm{n}}(x) = \mathcal{S}(x; \mathcal{Q}_{\mathrm{n}})$ is for normal task pairs. The ReLU function ensures that all values in the saliency map remain non-negative. This formulation isolates activations dominated by privacy-specific attention, yielding a spatially adaptive mask $\mathcal{A}(x)$ for subsequent perturbation allocation.

\paragraph{Dual-Task Adversarial Objective}
We employ a U-Net-style network $\mathcal{G}_\phi$ as the perturbation generator, which takes an input image $x$ and the contrastive attention map $\mathcal{A}(x)$ to produce a targeted perturbation. The objective is to generate noise that selectively shields privacy-sensitive regions while preserving overall task utility. To achieve this dual objective, we integrate the contrastive attention module into the generation process via affine modulation at each resolution level $k$ of the generator's decoder. Specifically, at level $k$, the attention map $\mathcal{A}(x)$ is downsampled to the current resolution, yielding $\mathcal{A}_k$. A lightweight convolutional network $\mathbf{L}_{\text{cnn}}^{(k)}$ (part of $\mathcal{G}_\phi$) takes $\mathcal{A}_k$ as input and predicts two modulation coefficient maps: $\gamma_k$ and $\beta_k$. These are mapped to bounded scale/shift coefficients:
\begin{equation}
    \tilde{\gamma}_k = 1 + s\,\gamma_k, \qquad
\tilde{\beta}_k = s\,\beta_k,
\end{equation}
where \(s\) is a fixed strength hyperparameter. Attention is injected by affine-modulating the \(k\)-th feature map \(\mathcal{F}_k\) to emphasize attended regions and add signal where attention is high:
\begin{equation}
    \hat{\mathcal{F}}_k = \tilde{\gamma}_k \odot \mathcal{F}_k + \tilde{\beta}_k,
\end{equation}
with \(\odot\) denoting element-wise multiplication. Let $\mathcal{G}'_\phi$ denote $\mathcal{G}_\phi$ the equipped with this injection at every layer; its output is squashed by \(\tanh\) to yield a per-pixel perturbation in \([-1,1]\):
\begin{equation}
\tilde{\delta}(x,\mathcal{A})=\varepsilon\, \mathcal{G}'_{\phi}(x;\mathcal{A}), \quad \|\delta(x,\mathcal{A})\|_{\infty}\le\varepsilon.
\end{equation}

To explicitly balance task utility and privacy protection through our novel dual-task mechanism, we define two complementary losses: the task-preservation loss $\mathcal{L}_{\mathrm{n}}(x_{\mathrm{adv}})$, which encourages high confidence in responses to non-private queries $\mathcal{Q}_n$ by minimizing negative log-likelihood, and the privacy-interference loss $\mathcal{L}_{\mathrm{p}}(x_{\mathrm{adv}})$, which penalizes fidelity on privacy-oriented ones:
\begin{equation}
    \left\{
    \begin{aligned}
        \mathcal{L}_{\mathrm{n}}(x_{\mathrm{adv}}) &= \sum_{(q,a) \in \mathcal{Q}_{\mathrm{n}}} - \log p_\theta(a \mid q, x_{\mathrm{adv}}), \\
        \mathcal{L}_{\mathrm{p}}(x_{\mathrm{adv}}) &= \sum_{(q,a) \in \mathcal{Q}_{\mathrm{p}}} - \log p_\theta(a \mid q, x_{\mathrm{adv}}).
    \end{aligned}
    \right.
\end{equation}
The generator $\mathcal{G}_\phi'$ is optimized via stochastic gradient descent to minimize the composite loss:
\begin{equation}
    \min_{\phi} \quad \alpha \mathbb{E} \big[ \mathcal{L}_{\mathrm{n}}(x_{\mathrm{adv}}) \big] - \beta \mathbb{E} \big[ \mathcal{L}_{\mathrm{p}}(x_{\mathrm{adv}}) \big],
\end{equation}
where $\alpha > 0$ and $\beta > 0$ empirically trades off task fidelity against leakage suppression. This dual-loss formulation ensures that minimizing $\mathcal{L}_{\mathrm{n}}$ preserves agent performance on normal tasks $\mathcal{Q}_{\mathrm{n}}$, while minimizing $\mathcal{L}_{\mathrm{p}}$ actively suppresses confidence on privacy extractions $\mathcal{Q}_{\mathrm{p}}$ (as $\log p_\theta < 0$ for $p_\theta < 1$). By integrating the attention-guided modulation, it achieves decoupled task control, ultimately yielding state-of-the-art privacy protection.

\section{Experiments}
\subsection{Experimental Setup}
\textbf{Baselines.} We utilize AnyAttack~\cite{zhang2025anyattack} FOA-Attack~\cite{jia2025adversarial} and VIP~\cite{meftah2025vipvisualinformationprotection} as baseline defenses for evaluating our framework by repurposing these state-of-the-art adversarial perturbation techniques for privacy protection in MLLMs. Specifically, for AnyAttack and FOA-Attack, we treat the clean image as a screenshot containing sensitive privacy information and the target image as a non-sensitive screenshot. We generate adversarial perturbations to obscure privacy leaks. For VIP, first mark the text locations in the image and rewrite the optimization target question as \textit{‘What is the privacy information in the image?’}. Additionally, we include an original baseline representing the undefended scenario, where clean screenshots are processed directly without any added noise.


\begin{table*}[ht]

\small
\centering
\caption{Main experimental results comparing four baseline methods across seven metrics on 2 Open-Source MLLMs (O-MLLMs), 2 Commercial MLLMs (C-MLLMs), and 2 Specialized GUI Agents. Best results are in \textbf{bold}. $\uparrow$ indicates metrics where higher values are better; $\downarrow$ indicates metrics where lower values are better.}
\label{Table1} 

\begin{adjustbox}{max width=\textwidth}

\begin{tabular}{l  l l c c c c c c c}
\toprule
 & Model & Attack Method & Acc $\uparrow$ & LR $\downarrow$ & MS $\downarrow$ & BertScore $\downarrow$ & CS $\downarrow$ & BLEU $\downarrow$ & ROUGE-L $\downarrow$ \\
\midrule

\multirow{10}{*}{\rotatebox[origin=c]{90}{\textbf{O-MLLMs}}}
& \multirow{5}{*}{InternVL3-5-8B}
 & Original & 83.00 & 96.19 & 92.96 & 0.7172 & 0.8866 & 0.5586 & 0.7770\\
 & & AnyAttack & 78.00 & 90.48 & 86.38 & 0.5689 & 0.7671 & 0.3512 & 0.5702\\
 & & FOA-Attack & 70.00 & 78.57 & 78.54 & 0.5706 & 0.7198 & 0.4229 & 0.6012\\
 & & VIP & 80.00 & 85.31 & 83.52 & 0.5801 & 0.8089 & 0.4022 & 0.6273\\
 & & \textbf{Ours} & \textbf{83.00} & \textbf{24.29} & \textbf{38.16} & \textbf{0.1905} & \textbf{0.3996} & \textbf{0.0841} & \textbf{0.1275}\\
 \cmidrule(l){2-10}
& \multirow{5}{*}{Qwen2.5-VL-7B }
 & Original & 89.00 & 97.14 & 96.99 & 0.8675 & 0.9011 & 0.7830 & 0.8991\\
 & & AnyAttack & 70.00 & 95.24 & 91.86 & 0.6721 & 0.8705 & 0.5164 & 0.7317\\
 & & FOA-Attack & 73.00 & 83.81 & 82.84 & 0.6307 & 0.8150 & 0.5343 & 0.7309\\
 & & VIP & 81.00 & 94.76 & 92.67 & 0.7556 & 0.8960 & 0.5932 & 0.7856\\
 & & \textbf{Ours} & \textbf{88.00} & \textbf{32.86} & \textbf{38.82} & \textbf{0.1998} & \textbf{0.4202} & \textbf{0.0908} & \textbf{0.1660}\\
 \midrule 

\multirow{10}{*}{\rotatebox[origin=c]{90}{\textbf{C-MLLMs}}}
 & \multirow{5}{*}{GPT-5}
 & Original & 93.00 & 97.14 & 97.15 & 0.9260 & 0.9627 & 0.8584 & 0.9246\\
 & & AnyAttack & 90.00 & 91.43 & 87.49 & 0.6757 & 0.8490 & 0.4978 & 0.7446\\
 & & FOA-Attack & 80.00 & 80.95 & 79.72 & 0.6293 & 0.7532 & 0.4967 & 0.7004\\
 & & VIP & 86.00 & 80.00 & 79.33 & 0.6662 & 0.7715 & 0.5284 & 0.6846\\
 & & \textbf{Ours} & \textbf{91.00} & \textbf{23.33} & \textbf{24.43} & \textbf{0.1589} & \textbf{0.2379} & \textbf{0.0941} & \textbf{0.1628}\\
 \cmidrule(l){2-10}
 & \multirow{5}{*}{Gemini-2.0 flash}
 & Original & 87.00 & 96.67 & 96.72 & 0.8651 & 0.9574 & 0.8030 & 0.9302\\
 & & AnyAttack & 82.00 & 97.14 & 96.66 & 0.7993 & 0.9400 & 0.6960 & 0.9001\\
 & & FOA-Attack & 79.00 & 94.76 & 94.13 & 0.7401 & 0.9045 & 0.6362 & 0.8342\\
 & & VIP & 84.00 & 94.79 & 92.76 & 0.7311 & 0.8830 & 0.5746 & 0.7828\\
 & & \textbf{Ours} & \textbf{87.00} & \textbf{57.62} & \textbf{61.41} & \textbf{0.3805} & \textbf{0.5193} & \textbf{0.1686} & \textbf{0.2592}\\
 \midrule

\multirow{10}{*}{\rotatebox[origin=c]{90}{\textbf{GUI Agents}}}
 & \multirow{5}{*}{Holo1.5-7B}
 & Original & 70.00 & 94.76 & 94.07 & 0.8351 & 0.9036 & 0.7182 & 0.8232\\
 & & AnyAttack & 61.00 & 81.43 & 79.07 & 0.5711 & 0.7447 & 0.3886 & 0.5801\\
 & & FOA-Attack & 58.00 & 78.10 & 78.29 & 0.5590 & 0.7571 & 0.3964 & 0.5852\\
 & & VIP & 57.00 & 89.05 & 86.19 & 0.6559 & 0.8263 & 0.4633 & 0.6583\\
 & & \textbf{Ours} & \textbf{69.00} & \textbf{17.14} & \textbf{30.75} & \textbf{0.1327} & \textbf{0.3620} & \textbf{0.0302} & \textbf{0.0949}\\
 \cmidrule(l){2-10}
 & \multirow{5}{*}{UI-TARS-7B}
 & Original & 80.00 & 93.81 & 93.10 & 0.7846 & 0.9079 & 0.6721 & 0.8363\\
 & & AnyAttack & \textbf{76.00} & 81.43 & 80.11 & 0.5533 & 0.7615 & 0.3748 & 0.6057\\
 & & FOA-Attack & 64.00 & 76.67 & 75.60 & 0.5349 & 0.7187 & 0.4226 & 0.6178\\
 & & VIP & 66.00 & 89.05 & 85.83 & 0.6739 & 0.8288 & 0.5232 & 0.7223\\
 & & \textbf{Ours} & 67.00 & \textbf{34.76} & \textbf{38.35} & \textbf{0.2278} & \textbf{0.4535} & \textbf{0.1139} & \textbf{0.2034}\\
\bottomrule
\end{tabular}

\end{adjustbox}
\end{table*}

\begin{figure*}[ht]
    \centering
    \includegraphics[width=1.0\linewidth]{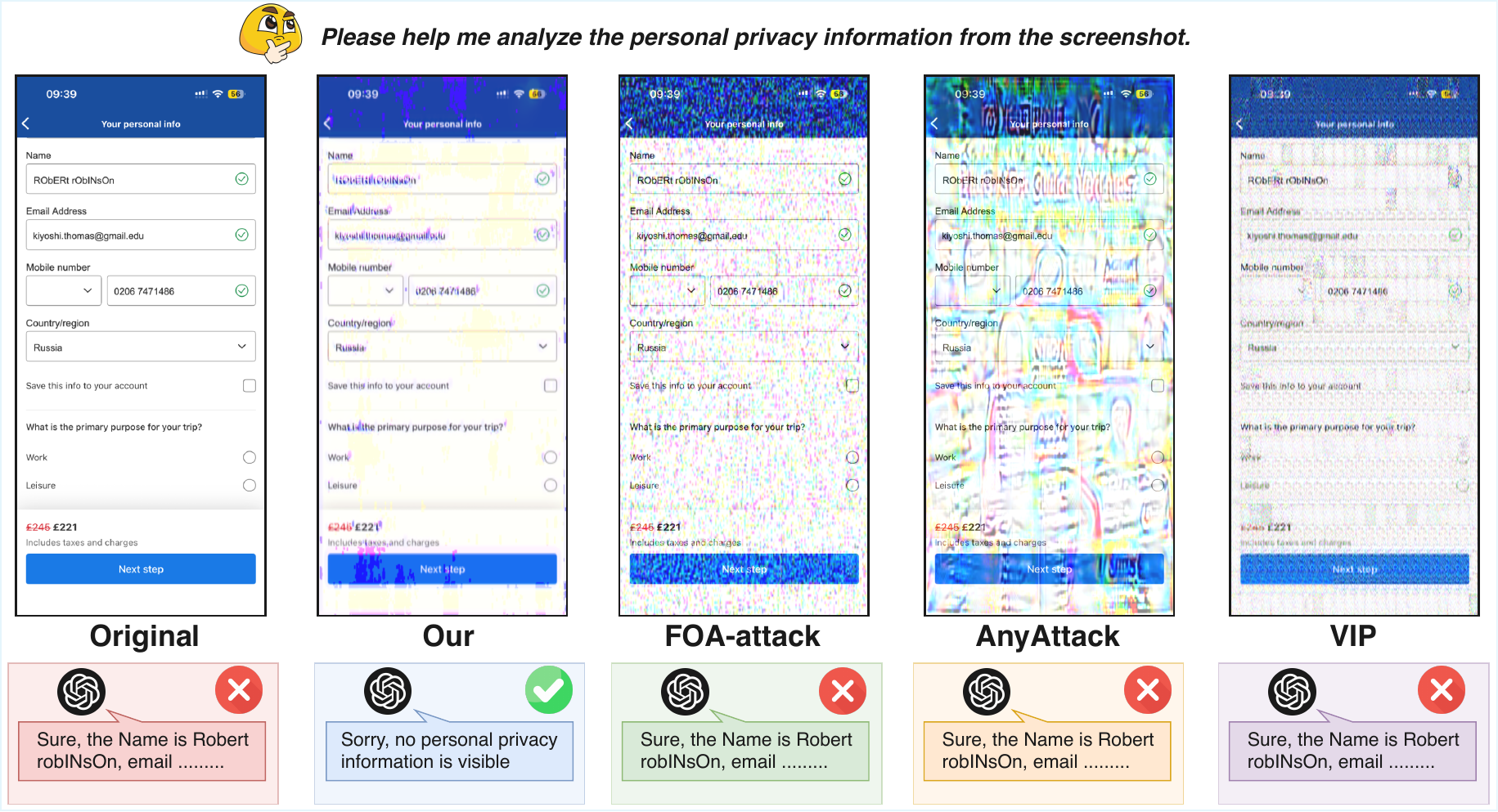}
    \caption{Comparison of adversarial images generated by different baselines after adding perturbations to the same privacy-sensitive image.}
    \label{differentfig}
\end{figure*}

\paragraph{Datasets and Models.}

This work introduces PrivScreen, a dual-task QA-style privacy protection dataset derived from real application screenshots. The dataset comprises over 500 screenshots, augmented with over 1000 synthetic PII, and includes two QA annotations per image: (i) a privacy-focused QA querying sensitive items (to assess leakage) and (ii) a utility-focused QA on general screen content (to evaluate functionality). Both QAs share the same screenshot to maintain identical visual distributions. An in-app split (80\% training, 20\% evaluation) minimizes cross-app bias and enhances generalization. We benchmark three model categories: (1) open-source MLLMs, including \textit{InternVL3}~\cite{zhu2025internvl3exploringadvancedtraining} and \textit{Qwen2.5}~\cite{qwen2.5}; (2) GUI-specialized MLLMs, such as \textit{Holo1.5}~\cite{hai2025holo15modelfamily} and \textit{UI-TARS}~\cite{qin2025ui}; and (3) commercial MLLMs, like \textit{GPT-5} and \textit{Gemini-2.0}. This diverse evaluation provides a robust benchmark across architectures and optimizations.

\paragraph{Metrics.}
To comprehensively evaluate our method and baselines, we assess normal task execution and privacy-protection efficacy metrics. For the normal task, which measures the agent’s ability to correctly complete intended operations on screenshot content. We report Accuracy (Acc) as the success rate. For the privacy information task—aimed at suppressing leakage of sensitive fields, we evaluate at two complementary levels: (i) character level, using Match Score (MS) for exact text matching, Leakage Rate (LR) defined as the proportion of samples with $\mathrm{MS}>0.6$, BLEU for $n$-gram precision with word-order sensitivity, and ROUGE-L for longest-common-subsequence similarity; and (ii) semantic level, using BERTScore to compare contextual embeddings and Cosine Similarity (CS) to measure vector-based semantic equivalence.

\paragraph{Implementation Details.}

In this work, supervision targets only the answer span, excluding question and visual placeholder tokens from the loss. Perturbations are generated using a U-Net-style network, with its output passed through a $\tanh$ activation and constrained by an $L_\infty$ norm ($\varepsilon = 128/255$) before being added to the original image and clipped to $[0,1]$. This level of perturbations aims to suppress PII extraction effectively while maintaining key UI semantics and spatial layout for robust GUI reasoning. The surrogate MLLM, \textit{InternVL3\_5-2B}, serves solely as the gradient target and remains frozen during training. Loss weights are $\alpha = \beta = 1.0$. Training uses Adam with a learning rate of $1 \times 10^{-4}$, batch size of 4, and 20 epochs. All experiments are run on a single NVIDIA L20 GPU.

\subsection{Main Results}


As shown in Table~\ref{Table1}, the Original setting confirms the severe privacy risks inherent in MLLM-based GUI agents. Across all six evaluated models, Original exhibits catastrophic leakage, with LR consistently above 93\% and the semantic metric BertScore indicating a high degree of information exposure. This establishes a critical need for robust protection. The experimental results mainly demonstrate the following key aspects:

\begin{itemize}
    \item \textbf{SOTA Privacy Protection.} Our \methodnorm framework achieves state-of-the-art privacy protection, outperforming all competing baselines (AnyAttack, FOA-Attack, VIP). This is evident in the results from SOTA OpenSource models. For instance, on \textit{Qwen-VL-7B-Instruct}, \methodnorm reduces the LR to 32.86\%, a massive reduction from the Original~(97.14\%) and superior to the baseline FOA-Attack~(83.81\%). This robust protection extends beyond simple keyword leakage to the semantic level: \methodnorm 's BertScore of 0.1998 is approximately 4.3$\times$ lower than the Original 0.8675 and 3.8$\times$ lower than VIP 0.7556, demonstrating a profound reduction in semantic information leakage.

    \item \textbf{Task Utility.} Crucially, \methodnorm achieves this exceptional privacy protection without sacrificing task utility. It consistently maintains the highest Acc among all defense methods, performing on par with the unprotected Original model. For example, on \textit{Qwen-VL-7B}, \methodnorm's accuracy~(88.00\%) is higher than all other defenses. On \textit{InternVL-5-8B}, \methodnorm achieves 83.00\% and even matches the Original accuracy exactly. This demonstrates \methodnorm's ability to effectively decouple the objectives of privacy protection and task preservation, resolving the critical privacy-utility trade-off.

    \item \textbf{Transferability.} To explore the transferability of \methodnorm, we tested it on six SOTA models across three types (Open Source, Specialized GUI, and Commercial). The framework's robustness is evident across different architectures. 
    For instance, on the commercial model \textit{GPT-5}, \method cuts the LR to a mere 23.33\% and reduces the BertScore to 0.1589, a 5.8$\times$ reduction from the Original's 0.9260. \method consistently establishes the best privacy protection and highest task utility (Acc) across all evaluated models, proving its effective transferability.
\end{itemize}

The results highlight the superiority of our dual-task adversarial objective and contrastive attention module. As shown in Figure~\ref{differentfig}, this design targets adversarial perturbations to privacy-sensitive regions, effectively decoupling privacy interference from task preservation. Overall, experiments confirm our method achieves state-of-the-art privacy protection across diverse MLLM agents while maintaining strong practical utility.

\begin{figure*}[ht]
    \centering
    \includegraphics[width=0.90\linewidth]{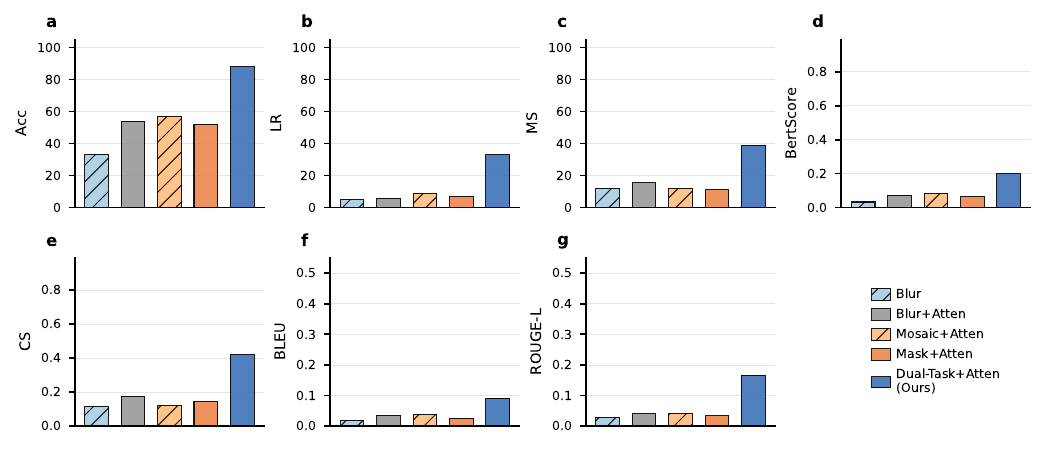}
    \caption{Comparison of the performance of five different module combinations on the Mobile GUI Agent.}
    \label{ablation}
\end{figure*}

\subsection{Ablation Study}

\paragraph{The Role of Attention.} As shown in Figure~\ref{ablation}~(a), when only Blur is used, the accuracy on normal tasks is only about 33\%, indicating that simple global blurring severely damages task-relevant key information. After introducing attention (Blur+Atten), the accuracy increases to about 54\%, which demonstrates that concentrating the perturbation more on privacy regions can significantly reduce interference with normal tasks while maintaining a similar level of privacy strength. However, because the MLLM still encodes the entire image in a unified manner, attention cannot be perfectly precise, and a certain amount of perturbation inevitably remains in non-privacy regions. For strong occlusion methods such as mosaic and mask, which directly overwrite pixels, the perturbation is more likely to spill over into task-relevant areas, causing a catastrophic impact on normal-task performance. Among them, Mask+Atten achieves only 52\% accuracy on normal tasks, even though it provides more thorough privacy masking. 

\begin{figure}
    \centering
    \includegraphics[width=1.0\linewidth]{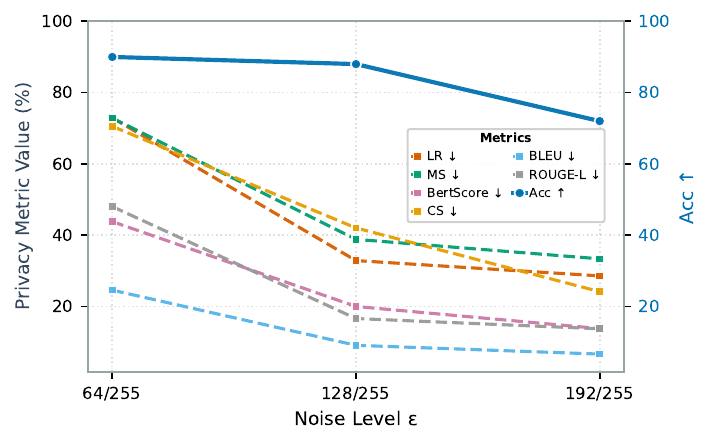}
    \caption{Impact of the different perturbation bounds}
    \label{differentbounds}
\end{figure}

\paragraph{The Importance of the Dual-Task Method.} In contrast, our proposed Dual-Task+Atten method exhibits a more balanced performance across all metrics. On the one hand, as shown in Figure~\ref{ablation}~(f)(g), all privacy-protection scores are below 0.2, meaning that at the semantic level, the text is no longer sufficient to reconstruct private content, indicating that the privacy protection effect is not weakened. On the other hand, our method achieves over 80\% accuracy on normal tasks. By jointly optimizing the objectives of privacy protection and task preservation during training, our method effectively constrains the spread of perturbations into normal-task regions and, while maintaining strong privacy protection, significantly alleviates the performance degradation on normal tasks.

\paragraph{Impact of the perturbation bound $\varepsilon$.}
As shown in Figure~\ref{differentbounds}, to evaluate the impact of noise levels $\varepsilon$ at 64/255, 128/255, and 192/255, experiments show that escalating $\varepsilon$ bolsters privacy by reducing privacy metrics from $80\sim100\%$ to $20\sim40\%$, yet it impairs task accuracy Acc from 88\% to 72\%. The primary setup employs $\varepsilon$=128/255, as it offers near-equivalent accuracy to 64/255 with just a slight $5\sim10\%$ reduction, while providing privacy protection almost on par with 192/255, gaining little more at the expense of greater utility degradation from intensified noise.

\begin{figure*}[ht]
    \centering
    \includegraphics[width=0.95\linewidth]{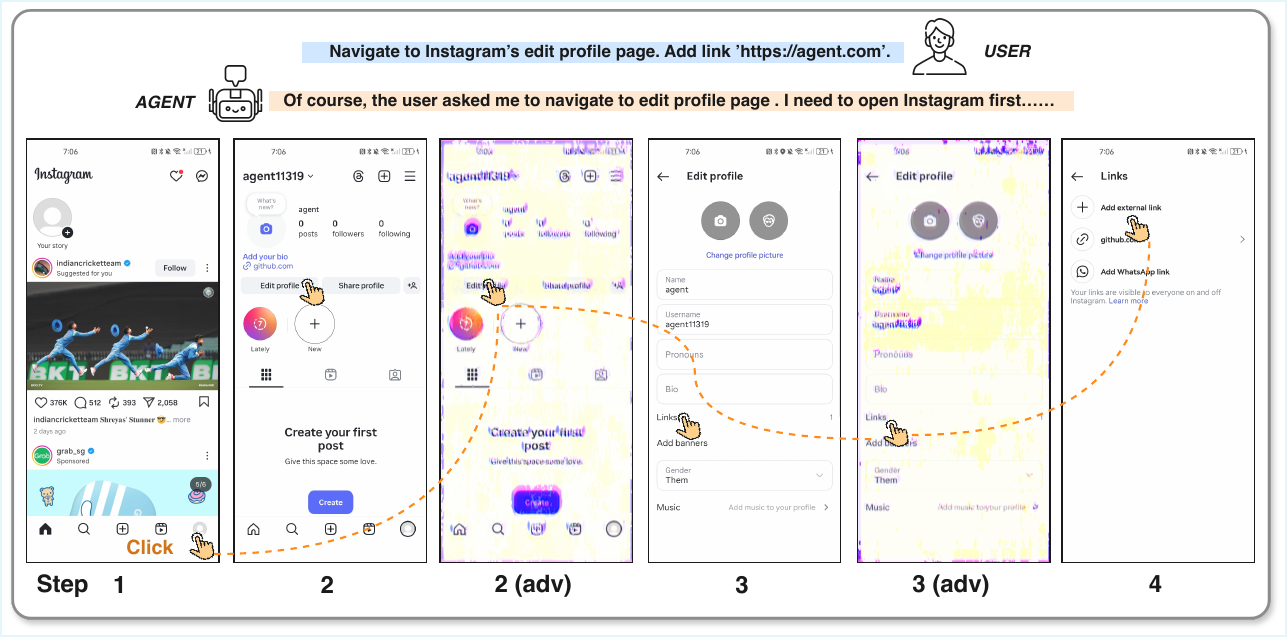}
    \caption{The task execution process on Mobile-Agent-V3, shown as a full sequence of screenshots.}
    \label{fig:realworldcase}
\end{figure*}

\subsection{Efficiency Comparison}

\begin{table}[ht]
\centering
\caption{Comparison of different privacy protection methods.}
\label{efficient}
\renewcommand\arraystretch{1.2}
\resizebox{\linewidth}{!}{
\begin{tabular}{lcccc}
\toprule
\textbf{Property} & \textbf{Ours} & \textbf{Anyattack} & \textbf{FOA-Attack} & \textbf{VIP} \\
\midrule
End-to-end pipeline        & \CheckmarkBold & \CheckmarkBold & \XSolidBrush  & \XSolidBrush  \\
No additional pre-training & \CheckmarkBold & \XSolidBrush   & \CheckmarkBold & \CheckmarkBold \\
Inference time / image     & $<0.3\,\mathrm{s}$ & $<0.3\,\mathrm{s}$ & $\approx120\,\mathrm{s}$ & $\approx600\,\mathrm{s}$ \\
\bottomrule
\end{tabular}
}
\end{table}
Table~\ref{efficient} compares the computational efficiency and simplicity of different baselines. The proposed method demonstrates clear advantages in both end-to-end integration and real-time inference efficiency. Unlike FOA-Attack and VIP, which rely on multi-stage optimization and extensive offline pre-training, \methodnorm operates as a fully end-to-end pipeline without additional pre-training, achieving inference times under 0.3$\mathrm{s}$ per image, comparable to lightweight methods like Anyattack but with a more stable and task-preserving design. In contrast, FOA-Attack and VIP require approximately 120$\mathrm{s}$ and 600$\mathrm{s}$ per image, making them impractical for real-time or mobile deployment. These results highlight that \methodnorm achieves an optimal trade-off between privacy protection and computational cost, allowing efficient execution on edge devices or within mobile MLLM agents. The low latency and independence from pre-training make it particularly suitable for on-device inference scenarios, where energy efficiency, responsiveness, and user privacy must be simultaneously preserved.




\subsection{Real-World Case}
We evaluated our approach through real-device task execution on a HarmonyOS smartphone using Mobile-Agent-V3~\cite{ye2025mobile}. Across a range of representative interaction workflows, the agent successfully perceives the interface, issues actions, and completes tasks end-to-end with our privacy protection module enabled. As illustrated in Figure~\ref{fig:realworldcase}, even after applying our protection mechanisms, the tasks continue to function normally, with no noticeable degradation in success rate or interaction quality. Moreover, our generator is lightweight and plug-and-play, which is approximately 300 MB, making it feasible for on-device deployment on commodity smartphones. Taken together, these findings indicate that our method can be deployed on real devices while preserving both privacy protection and the agent’s practical usability.

\section{Conclusion}
In this work, we reveal and address a critical privacy leakage risk inherent in mobile MLLM agents: external API routers can exploit MLLMs to extract private information from user screenshots. Existing privacy protection methods fail to balance task performance and privacy protection. To tackle this, we propose \method, an adversarial protector that innovatively integrates a contrastive attention module focused on sensitive regions into the generator and, for the first time, introduces a dual-task adversarial objective, enabling explicit disentanglement of normal and privacy tasks. By jointly optimizing task-preservation and privacy-interference losses, \method effectively suppresses PII exposure without compromising the agent’s functionality. To further support this goal, we contribute PrivScreen, a multi-application mobile screenshot dataset for privacy and interface understanding. Experimental results across 6 MLLMs demonstrate the significant advantage of our method, substantially reducing privacy leakage while maintaining normal task execution. This work establishes a new paradigm for privacy protection in MLLM agents, shifting the research focus from traditional perception-level anomaly detection to adversarial disentanglement of task utility and privacy information at the inference level.


\clearpage

{
    \small
    \bibliographystyle{ieeenat_fullname}
    \bibliography{main}
}

\end{document}